\newcommand{\rem}[1]{}

\documentclass[prb,twocolumn,showpacs,amsmath,amssymb]{revtex4}
\usepackage{epsfig}
\usepackage{amsmath,graphicx,subfigure}

\begin{document}

\title{Non-Hermitian Spectra and Anderson Localization}
\author{Luca G. Molinari}
\affiliation{Dipartimento di Fisica and I.N.F.N. sezione di Milano\\
Via Celoria 16, 20133 Milano, Italy\\
E-mail: luca.molinari@mi.infn.it}
\date{october 2008}
\begin{abstract}
The spectrum of exponents of the transfer matrix provides the localization 
lengths of Anderson's model for a particle in a lattice with disordered 
potential. 
I show that a duality identity for determinants and Jensen's identity for 
subharmonic functions, give a formula for the spectrum in terms of 
eigenvalues of the Hamiltonian with non-Hermitian boundary conditions.
The formula is exact; it involves an average over a Bloch phase, rather 
than disorder.
A preliminary investigation of non-Hermitian spectra of Anderson's model in 
D=1,2 and on the smallest exponent is presented.
\end{abstract}
\maketitle

\section{Introduction}
Several models in physics are described by matrices with banded 
or block-tridiagonal structure. Examples are the Laplacian matrix, the 
Anderson Hamiltonian for transport in a lattice with random impurities, 
band random matrices, tight binding models in condensed matter and chemistry. 
The matrix 
structure reproduces that of a system consisting of a chain of units with same 
number of internal states, with nearest neighbors interaction. 
Finite size effects
are often dealt with by imposing periodicity; the limit of large number of 
units is eventually taken.

The matrix structure calls for a transfer matrix description of the 
eigenstates, and the spectrum of exponents of the 
transfer matrix describes the decay lengths of the eigenstates. 
For the Anderson model or band random matrices, most of the knowledge 
on Lyapunov spectra relies on numerical computations.

With great generality, I showed that an analytic tool to access the decay 
lengths is a {\em duality relation}, that connects the spectrum of a 
non-Hermitian extension of the block-tridiagonal matrix, with the spectrum 
of the related transfer 
matrix\cite{Molinari97,Molinari98,Molinari03,Molinari07}.
The extension arises by mere generalization of boundary conditions for the
eigenstates. If $\vec u_k$ specifies the state of a unit of the chain
($k=1,\ldots ,n$), the 
boundary conditions (b.c.) are parametrized by a complex number $z$:
\begin{eqnarray}
\vec u_{n+1}=z^n\vec u_1, \quad \vec u_0=\frac{1}{z^n}\vec u_n
\label{generalized}
\end{eqnarray}
This destroys Hermiticity of the Hamiltonian matrix, but enlights a nice
property of the transfer matrix:
\begin{eqnarray}
T(\epsilon )\left[\begin{array}c
 \vec u_1 \\ \frac{1}{z^n}\vec u_n
\end{array}\right] = z^n
\left[\begin{array}c
 \vec u_1 \\ \frac{1}{z^n}\vec u_n\end{array}\right] \label{eigenvT}
\end{eqnarray}
The ensuing spectral duality and Jensen's identity for subharmonic functions 
allow to evaluate the counting function of exponents.
This paper is intended to introduce the theory and explore its
application to the long-studied problem of Anderson's localization.
In section II the duality relation is reviewed and the main formula 
(\ref{spectrum}) for the exponents is obtained from Jensen's theorem.
The theory can be extended to include the spectrum of the (Lyapunov)
exponents of the matrix T$^\dagger $T, by constructing a corresponding 
non-Hermitian block tridiagonal matrix, twice the size of the original 
Hamiltonian matrix. In section III a preliminary study of the eigenvalues 
of non Hermitian Hamiltonian matrices in D=1 and D=2 is made, with the 
purpose of illustrating the duality. The spectral formula is used to evaluate
the smallest exponent $\xi_{min}$, in a regime where the eigenvalues are
already all complex.

Because of their relevance in mathematics, numerical analysis and physics, 
block tridiagonal matrices are an active area of 
research\cite{Nabben99,Korotyaev07,Baldes08}. This work extends in
a new perspective the work by Hatano and Nelson\cite{Hatano96} which,
together with the works by Feinberg and Zee\cite{Feinberg97}, started an 
interest for non-Hermitian matrix models in physics.

\section{Theory}
\subsection{Transfer Matrix}
Consider the following block tridiagonal matrix with corners, of size 
$nm\times nm$,
\begin{eqnarray}
H=\left [\begin{array}{cccc}
A_1 &    B_1   &   {}  & B_n^\dagger \\
B_1^\dagger & \ddots &\ddots   & {}    \\
{}  & \ddots & \ddots  & B_{n-1}   \\
B_n &    {}  &   B^\dagger_{n-1}   & A_n   
\end{array}\right ]
\label{matrix}
\end{eqnarray}
The blocks have size $m\times m$: $B_k$ are complex matrices with 
$\det B_k\neq 0$, $A_k$ are Hermitian matrices. To the matrix 
$\epsilon I_{nm}-H$ 
there corresponds the {\em transfer matrix}\cite{Molinari97}
\begin{eqnarray}
&&T(\epsilon) = 
\left [\begin{array}{cc} B_n^{-1}(\epsilon I_m-A_n) & -B_n^{-1}B_{n-1}^\dagger
\\ I_m & 0 \end{array}\right ]\times \nonumber\\
&&\cdots \times
\left [\begin{array}{cc} B_1^{-1}(\epsilon I_m-A_1) & -B_1^{-1}B_n^\dagger
\\ I_m & 0 \end{array}\right ].
\end{eqnarray}
$I_m$ is the $m\times m$ identity matrix. 
The transfer matrix is so named because it transforms 
the eigenvalue equation $Hu=\epsilon u$ into a relation for 
the end-components of the vector $u=$ $(\vec u_1,\ldots ,\vec u_n)^t$:
\begin{eqnarray}
T(\epsilon )\left[\begin{array}c
 \vec u_1 \\ \vec u_0
\end{array}\right] = 
\left[\begin{array}c
 \vec u_{n+1} \\ \vec u_n\end{array}\right] \label{transferT}
\end{eqnarray}
The corners imply a condition of periodicity 
$\vec u_0=\vec u_n$ and $\vec u_{n+1}=\vec u_1$, that 
can be used to obtain the eigenvalue $\epsilon $ 
in alternative to diagonalization of $H$. 
By comparing  
eqs.(\ref{eigenvT}) and (\ref{transferT}) one arrives at the main point:
{\em to study the spectrum of}\,  $T(\epsilon )$, 
{\em one must impose the generalized b.c.} (\ref{generalized}). 
We thus introduce an instrumental non-Hermitian 
matrix depending on a parameter $z$ ($0\le {\rm arg}z\le \frac{2\pi}{n}$) 
\begin{eqnarray}
H(z^n)=\left [\begin{array}{cccc}
A_1 &    B_1   &   {}  & \frac{1}{z^n}B_n^\dagger \\
B_1^\dagger & \ddots &\ddots   & {}    \\
{}  & \ddots & \ddots  & B_{n-1}   \\
z^n B_n &    {}  &   B^\dagger_{n-1}   & A_n   \\
\end{array}\right ].
\label{nHmatrix}
\end{eqnarray}
The matrix 
is Hermitian for Bloch b.c. ($|z|=1$) but, for the purpose of studying
the spectrum of $T(\epsilon )$, it will be considered for $z\in C_0$.
The matrix can be brought by similarity to the balanced form 
$H_b(z)=Z^{-1}H(z^n)Z$,
\begin{eqnarray}
H_b(z)= \left [\begin{array}{cccc}
A_1            &    zB_1   &   {}    & \frac{1}{z}B_n^\dagger \\
\frac{1}{z}B_1^\dagger & \ddots    &  \ddots &     {}         \\
{}             & \ddots    &  \ddots & zB_{n-1}           \\
 zB_n          &    {}     & \frac{1}{z}B_{n-1}^\dagger  & A_n \\
\end{array}\right ],\label{balanced}
\end{eqnarray}
by means of the block diagonal matrix $Z$ with blocks $\{zI_m, \ldots 
,z^nI_m \}$.
Therefore, no site of the chain is privileged. 
While the matrix $H_b(z)$ does change if arg $z$ is increased by ${2\pi}/n$, 
its eigenvalues do not.


\subsection{Symplectic properties and exponents}
The following relations hold for the transfer matrix:
\begin{eqnarray}
&&T(\epsilon^*)^\dagger \Sigma_n T(\epsilon ) = \Sigma_n , \qquad 
\Sigma_n = \left[\begin{array}{cc} 0&-B_n^\dagger \\ B_n & 0
\end{array}\right]  \label{sigma}\\
&&T(\epsilon )\Sigma_n^{-1} T(\epsilon^*)^\dagger = \Sigma_n^{-1}, \qquad 
\Sigma_n^{-1} = \left[\begin{array}{cc} 0& B_n^{-1} \\ -{B_n^\dagger}^{-1} & 0
\end{array}\right]  \nonumber
\end{eqnarray}
Let us denote as $z_1^n\ldots z_{2m}^n$ the $2m$ eigenvalues of $T(\epsilon )$.
The relations imply that if $z_a^n$ is an eigenvalue of $T(\epsilon)$, 
then $(z_a^{-n})^*$ is an eigenvalue of $T(\epsilon^*)$.
In this study we are concerned with the {\em exponents} 
\begin{equation}
\xi_a (\epsilon ) = \log |z_a| \label{exponent}
\end{equation}
In general they still depend on $n$. Since $|\det T(\epsilon )|=1$, it is
always $\sum_a \xi_a(\epsilon )=0$. For real $\epsilon $ the exponents 
of $T(\epsilon )$ come in pairs $\pm\xi_a$.

\subsection{Duality, Jensen, and spectrum of exponents}
Since the extremal components $\vec u_1$ and $\vec u_n$ of the eigenvector 
$H(z^n) u = \epsilon u$ enter in the eigenvalue equation (\ref{eigenvT})
of $T(\epsilon)$, it follows that the characteristic polynomials of the two 
matrices are linked by a
\vskip0.5truecm
 
{\bf Duality relation}.\,
{\em $\epsilon $ is an eigenvalue of $H(z^n)$ iff $z^n$ is eigenvalue of 
$T(\epsilon )$}:
\begin{equation}
\frac{\det [\epsilon I_{nm} -H(z^n)]}{\det[B_1\cdots B_n]} 
= \frac{(-1)^{m}}{z^{nm}} \det [T(\epsilon )- z^n I_{2m}] 
\label{duality}
\end{equation} 
A proof of duality that holds also for non-Hermitian matrices, 
with blocks $B_k^\dagger $ being replaced by blocks $C_k$,
is found in ref.\cite{Molinari07}. 

The spectrum of exponents can be 
obtained from the spectrum of $H(z^n)$
through the following identity for analytic functions, 
which is a particular case of a theorem by Poisson and Jensen for 
subharmonic functions\cite{Markushevich}: 
\vskip0.5truecm

{\bf Jensen's identity}.\, 
{\em Let $f$ be an analytic function in the open disk of radius $R$, where it has zeros $z_1,\ldots , z_k$ that are ordered according to increasing modulus. Then, if $0<|z_1|$ and for $r$ such that $|z_\ell|\le r\le |z_{\ell +1}|$ we have:}
\begin{eqnarray}
\int_0^{2\pi} \frac{d\varphi}{2\pi} \log |f(re^{i\varphi})| = 
\log \frac{r^\ell |f(0)|}{|z_1\cdots z_\ell|} 
\end{eqnarray} 
\vskip0.5truecm

{\bf Proposition}.\, {\em For real $\xi$ and complex $\epsilon $ it is}
\begin{eqnarray}
&&\frac{1}{m}\sum_{\xi_a<\xi} [\xi-\xi_a(\epsilon) ]-\xi\,=\,
-\frac{1}{nm}\sum_{k=1}^n\log |\det B_k|\nonumber\\ 
&& +\int_0^{2\pi} \frac{d\varphi}{2\pi} \frac{1}{nm}\log |\det [\epsilon I_{nm}
-H (e^{n\xi+i\varphi})] |\label{spectrum} 
\end{eqnarray}
Proof: Jensen's identity is applied to the polynomial 
$f(z)=\det[T(\epsilon )-z^n I_{2m}]$, with $|f(0)|=1$ and 
$z=e^{\xi+i\varphi/n}$. The duality relation is then used to 
obtain the formula. $\blacksquare$

For $\xi =0$ a formula for the sum of positive
exponents follows. It involves a real eigenvalue spectrum
\begin{eqnarray}
&&\frac{1}{m}\sum_{\xi_a>0} \xi_a (\epsilon ) = 
-\frac{1}{nm}\sum_k\log |\det B_k|\nonumber \\
&& +\int_0^{2\pi} \frac{d\varphi}{2\pi} \frac{1}{nm}
\log |\det [\epsilon I_{nm}-H(e^{i\varphi})]\label{sumofexps}.
\end{eqnarray}

Equations (\ref{spectrum}) and (\ref{sumofexps}) are exact and 
valid for a single, general transfer matrix. In the theory of disordered 
systems, a formula for the sum of exponents is known, where Jensen's angular 
average is replaced by the ensemble 
average\cite{Lacroix86,PasturFigotin,Derrida00}.

\begin{figure}
\begin{center}
\includegraphics[width=4cm]{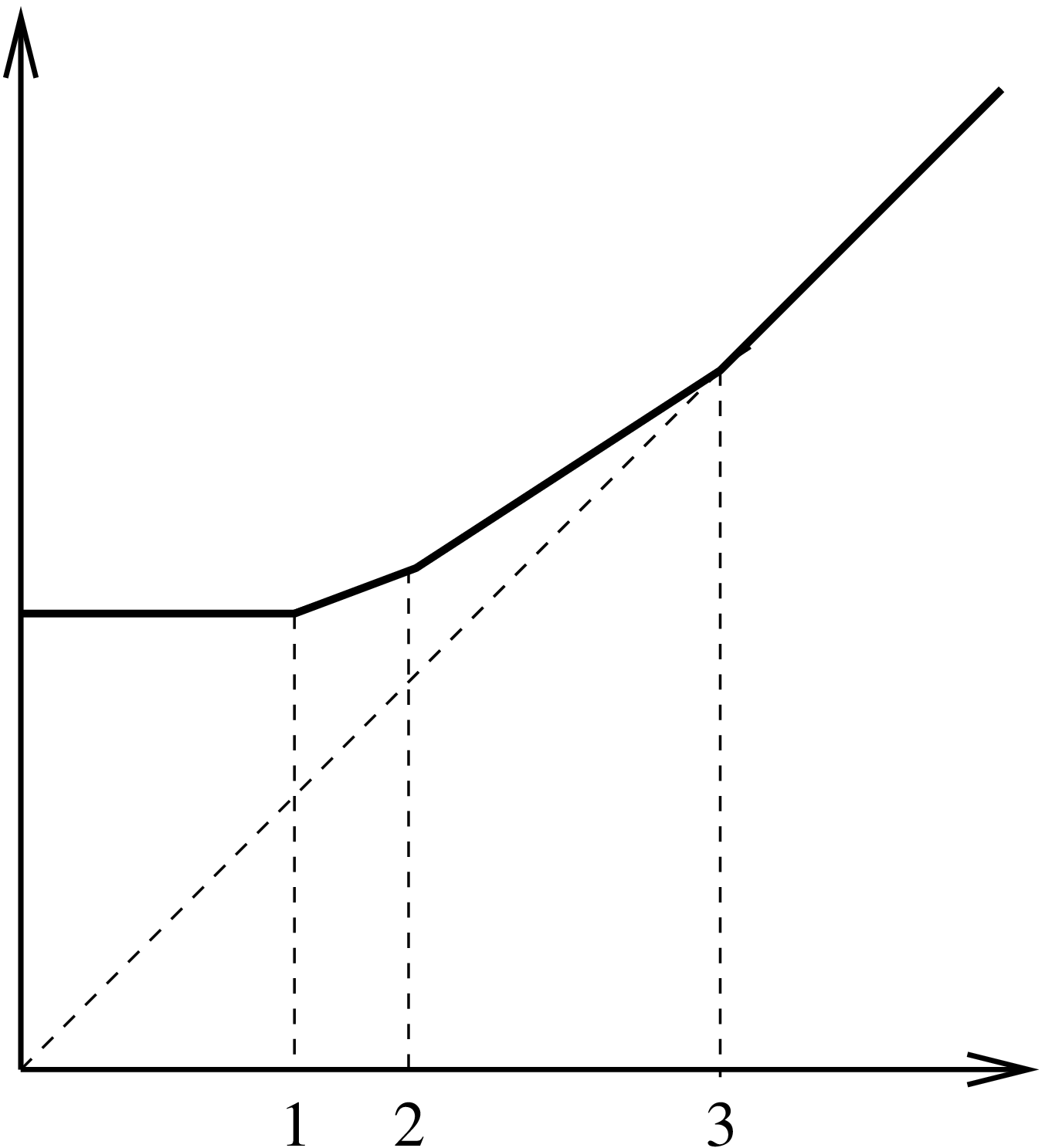}
\caption{\label{xiplot} The behaviour of the right hand side of 
eq.(\ref{spectrum}) as a function of $\xi$, for $m=3$. The constant value 
for $\xi<\xi_1=\xi_{min}$ is $\frac{1}{3}(\xi_1+\xi_2+\xi_3)$. 
At $\xi=\xi_1,\xi_2$ and $\xi_3=\xi_{MAX}$ the slope increases by $1/3$.}
\end{center}
\end{figure}

The left-hand side of eq.(\ref{spectrum}) is a non decreasing function 
of $\xi$ (Fig.\ref{xiplot}). 
For all $\xi\ge \xi_{MAX}(\epsilon)$ (the maximum exponent of 
$T(\epsilon)$), the right side is always equal to $\xi $. For all positive 
$\xi <\xi_{min}(\epsilon) $ (the smallest positive exponent), the right hand 
side is constant and equal to the average value of the exponents 
(\ref{sumofexps}).
For intermediate positive values of $\xi $ the function is piecewise linear, 
with discontinuities of order $1/m$ in the first derivative, at the values of 
the exponents.
\subsection{The matrix T$^\dagger$T}
Let us introduce the matrix $Q(\epsilon) =T(\epsilon^*)^\dagger T(\epsilon)$, 
with exponents $\gamma_a (\epsilon)$. For real $\epsilon $ the matrix is real 
and positive, and is preferred to $T$ because of better large $n$ behaviour 
of the exponents. If $B_n B_n^\dagger =I_m$, the matrix $Q$ is symplectic 
\begin{equation}
Q(\epsilon )\Sigma_n Q(\epsilon )=\Sigma_n 
\label{symplectictilde}\end{equation}
and the exponents come in pairs $\pm\gamma_a$.
Hereafter, the matrix $B_n$ will be restricted to be unitary. 
Under this restriction, it is shown in Appendix A that $(-1)^n 
Q(\epsilon)$ is unitarily equivalent to the transfer matrix 
$\theta(\epsilon )$
\begin{equation}
Q(\epsilon ) = (-1)^n 
\left[\begin{array}{cc} I_m & 0\\ 0 & B_n\end{array}\right] 
\theta(\epsilon )
\left[\begin{array}{cc} I_m & 0\\ 0 & B_n^\dagger\end{array}\right]
\label{teta}
\end{equation}
of the block tridiagonal matrix $M(\epsilon):$ 
\begin{eqnarray}
\left [\begin{array}{cccccccc}
A_1-\epsilon       & B_1   &  {}    & {}     & {}&  {}   &  {}   & I_m \\
B_1^\dagger &\ddots &\ddots  & {}     & {}&  {}   &  {}   & {}  \\
{}          &\ddots &\ddots  & B_{n-1}& {}&  {}   &  {}   & {}  \\
{}          &{} &B_{n-1}^\dagger &A_n-\epsilon & I_m &  {}   &  {}   & {}  \\ 
{}  & {}    & {}& I_m  &\epsilon -A_n &  B_{n-1}^\dagger   &  {}   & {}  \\ 
{}  & {}    & {}    & {}   &  B_{n-1}  &\ddots &\ddots & {}  \\ 
{}  & {}    & {}    & {}   & {}    &\ddots &\ddots & B_1^\dagger   \\ 
I_m   & {}    & {}    & {}   & {}    &  {}   & B_1  &\epsilon -A_1\\ 
\end{array}\right ]\nonumber
\end{eqnarray} 
Factors $z^{\pm 2n}$ are then introduced in the corners 
as in eq.(\ref{matrix}) and, because of the factor $(-1)^n$ in (\ref{teta}), 
the duality relation is as follows:
\begin{equation}
\frac{\det M(\epsilon,z^{2n})}{\prod_k |\det B_k|^2}=  
\frac{(-1)^m}{z^{2nm}} \det [Q(\epsilon ) - (iz)^{2n}I_{2m}]
\nonumber
\end{equation}
Since only the determinant matters, there is freedom to modify $M$ 
to a form where $\epsilon $ enters as a shift.
Left and right multiplication by the block diagonal matrices 
$\{I_{nm},I_m,-I_m,+I_m,\ldots\}$
and $\{I_{nm},-I_m,I_m,-I_m,\ldots\}$ give 
$\det M = (-1)^{nm}
\det [K((iz)^{2n})-\epsilon I_{2nm}]$, with $K((iz)^{2n})=$
\begin{eqnarray}
\left [\begin{array}{cccccccc}
A_1 & B_1     &  {}   & {}   & {}    &  {}   &  {}   & \frac{1}{(iz)^{2n}} \\
B_1^\dagger &\ddots &\ddots & {}   & {}        &  {}   &  {}   & {}  \\
{}  &\ddots &\ddots & B_{n-1}   & {}           &  {}   &  {}   & {}  \\ 
{}  & {}    & B_{n-1}^\dagger &A_n & -I_m    &  {}   &  {}   & {}  \\ 
{}  & {}    & {}    & I_m  & A_n &B_{n-1}^\dagger  &  {}   & {}  \\ 
{}  & {}    & {}    & {}   & B_{n-1}           &\ddots &\ddots & {}  \\ 
{}  & {}    & {}    & {}   & {}    &\ddots &\ddots & B_1^\dagger     \\ 
-(iz)^{2n} & {}   & {}   & {}    & {}    &  {}   & B_1       & A_1\\
\end{array}\right ]\nonumber 
\end{eqnarray}
A true duality relation among eigenvalues is obtained
(replace $iz$ by $z$)
\begin{equation}
\frac{\det [K(z^{2n})-\epsilon I_{2nm}]}{\prod_k |\det B_k|^2}=
\frac{(-1)^m}{z^{2nm}} \det [Q(\epsilon )-z^{2n}I_{2m}]
\end{equation}
The Lyapunov spectrum is extracted by means of Jensen's formula:
\begin{eqnarray}
&&\xi +\frac{1}{m}\sum_{ \gamma_a>\xi}[ \gamma_a(\epsilon )-\xi] = 
\frac{1}{nm}\sum_{k=1}^{n-1}\log |\det B_k|
\nonumber \\
&&+\int_0^{2\pi}\frac{d\varphi}{2\pi}\frac{1}{2nm}
\log |\det [K(e^{2n\xi+i\varphi})-\epsilon I_{2nm}]|\label{lyapspectrum}
\end{eqnarray}
Some properties of the matrix $K$ are presented in Appendix B.

\section{THE ANDERSON MODEL}
The discrete Anderson model describes a particle in a 
lattice, subject to a random potential. The potential of a sample is 
specified by a set $\{v_{\bf j}\}$ of random numbers chosen 
independently. Anderson\cite{Anderson58} considered a uniform density 
$p(v)=1/w$ in the interval $[-w/2,w/2]$. Lloyd\cite{Lloyd69,Mudry98} 
studied the Cauchy distribution $p(v)=\frac{\delta}{\pi}(v^2+\delta^2)^{-1}$, 
and evaluated the energy distribution exactly in any space dimension.
Anderson's choice and the simple hypercubic geometry are here considered. 
More complex lattices can be studied by transfer matrix\cite{Eilmes08}.

For a given configuration of potential, the eigenvalue equation is 
\begin{eqnarray}
\sum_{\bf e}u_{{\bf j}+{\bf e}} + 
v_{\bf j}u_{\bf j}\, =\, \epsilon u_{\bf j}
\end{eqnarray}
The sum is on the unit vectors along the $2D$ directions, $\epsilon $ 
is the energy of the particle, the lattice has lengths $n_1, \ldots ,n_D$.
If the $D$ axis is singled out, the sample is viewed as a number $n\equiv n_D$ 
of sections each containing $m\equiv n_1\cdots n_{D-1}$ sites. 
Accordingly, the Hamiltonian matrix is block tridiagonal
\begin{eqnarray}
H=\left [\begin{array}{cccc}
A_1 &    I_m   &   {}  & I_m \\
I_m & \ddots &\ddots   & {}    \\
{}  & \ddots & \ddots  & I_m   \\
I_m &    {}  &  I_m   & A_n   
\end{array}\right ]
\label{andersonmatrix}
\end{eqnarray}
with Hermitian blocks $A_i$ describing sections, 
and off diagonal blocks describing hopping among sections. The associated
transfer matrix is 
\begin{eqnarray}
T(\epsilon ) = \prod_{j=1}^{n}\left [\begin{array}{cc} \epsilon I_m-A_j & -I_m
\\ I_m & 0\\ \end{array}\right ]
\end{eqnarray}
For large $n$ the exponents of $T(\epsilon )$ describe the inverse decay 
lengths of the eigenstates of Anderson's Hamiltonian. To study them, 
we introduce b.c. terms $\pm z^n$ in the corner blocks of 
(\ref{andersonmatrix}), and choose periodic b.c. in the other $D-1$ 
directions, that appear in the diagonal blocks.

\begin{figure}
\begin{center}
\includegraphics[width=4cm]{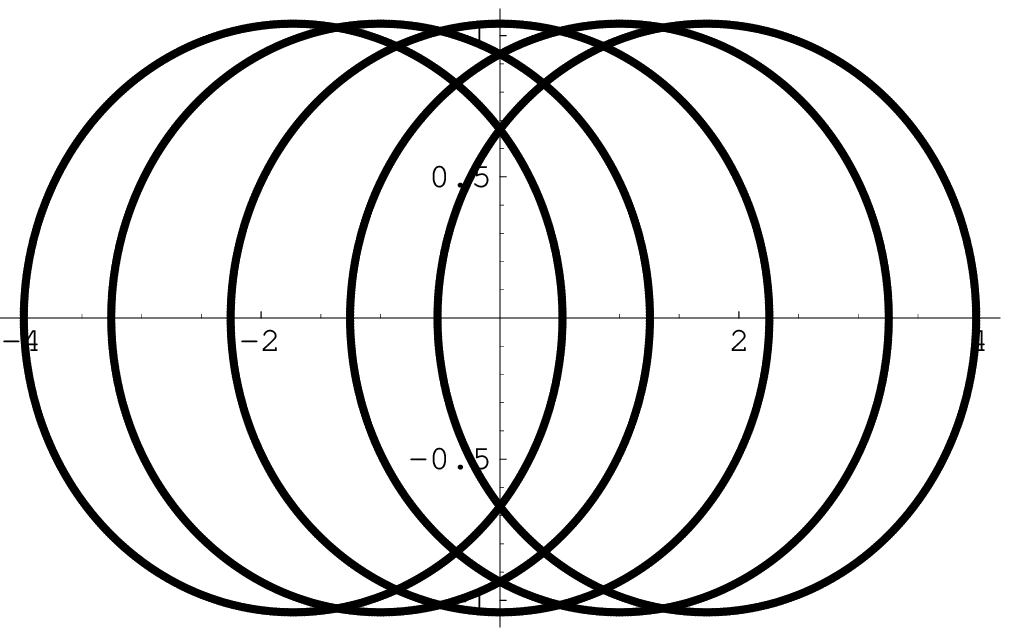}\hspace{0.4cm}
\includegraphics[width=4cm]{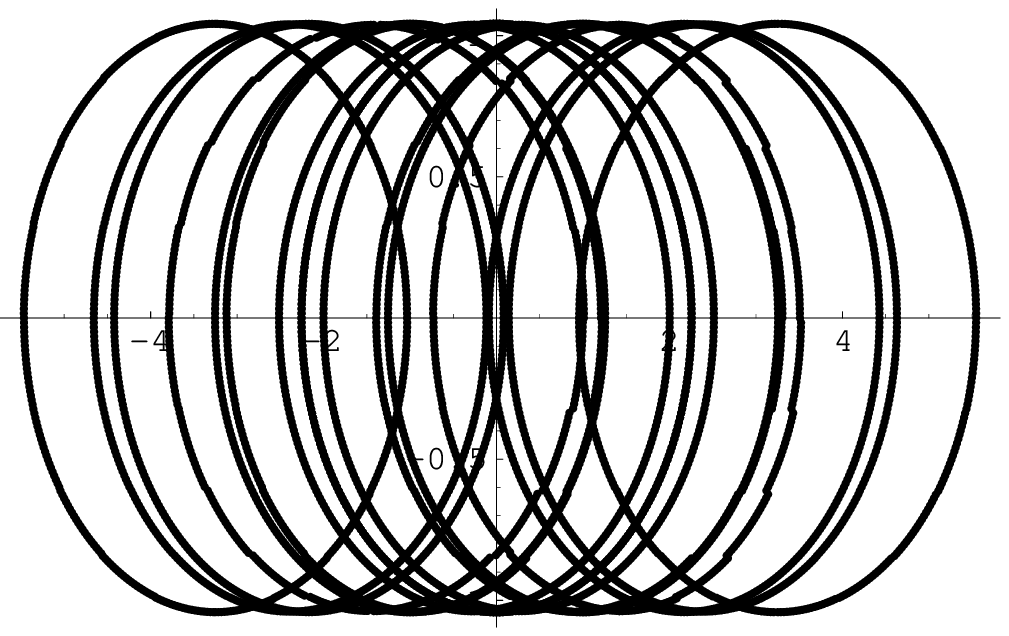}
\caption{\label{a0} Complex energy spectra for zero disorder, large $n$,
$\xi=1$: in 2D (left, $m=5$) and 3D (right, $n_x=n_y=4$ i.e. $m=16$).}
\end{center}
\end{figure}

{\em Remark 0}: For zero disorder the eigenvalues of $H(z^n)$ 
are complex for any nonzero value of the parameter $\xi $ that measures 
non-Hermiticity ($z=e^{\xi+i\varphi}$):
\begin{eqnarray}
&&{\rm Re}\, \epsilon =  2\cosh\xi \cos(\varphi+\frac{2\pi}{n}\ell)
+\epsilon_r \nonumber\\ 
&&{\rm Im}\, \epsilon =  2\sinh\xi \sin(\varphi+\frac{2\pi}{n}\ell),
\nonumber 
\end{eqnarray}
$\ell=1,\ldots ,n$ and  $r=1,\ldots, m$. There are $n$ eigenvalues 
on each ellipse centered at 
$ \epsilon_r = 2\sum_{i=1}^{D-1}\cos(2\pi \frac{k_i}{n_i})$, $1\le k_i\le n_i$.
Therefore, the spectrum has support on $m$ identical but shifted ellipses.
In $1D$ there is a single ellipse centered in the origin. In $2D$
there are $m=n_x$ distinct ones, while in $3D$ some of the $m=n_xn_y$ ellipses 
may overlap because centers may be degenerate (Fig.\ref{a0}). 
In Appendix C it is shown that the exponents of $T$ and 
$T^\dagger T$ coincide, for large $n$. 

{\em Remark 1}: 
For non-zero disorder the eigenvalues of the Hamiltonian matrix $H(z^n)$ 
are all contained inside the union of ellipses
\begin{eqnarray}
\frac{({\rm Re}\epsilon-\epsilon_0)^2}{4\cosh^2 \xi}+
\frac{({\rm Im}\epsilon)^2}{4\sinh^2\xi}
\le 1
\end{eqnarray}
where $\epsilon_0$ ranges in the interval $[-2D+2-w/2,2D-2+w/2]$.
Proof: If $H_b(z)u=\epsilon u$, and $u$ is normalized, the inner product 
$\epsilon =(u|H_b(z)u)$ in $C^{nm}$ is separated into real and imaginary parts:
\begin{eqnarray}
 &&{\rm Re}\epsilon -(u|Au) = 2|(u|Su)|\,\cosh \xi \, \cos(\varphi+\theta)
\nonumber\\
 && {\rm Im}\epsilon =2|(u|Su)|\,\sinh\xi \,\sin(\varphi+\theta)\nonumber
\end{eqnarray}
$A$ is the block diagonal part of $H_b$, $S$ is the one-block shift 
matrix, and $\theta =$arg$ (u|Su)$. The real number $(u|Au)=\epsilon_0$ ranges 
in the spectrum of $A$. Schwartz's inequality gives the bounds. 

{\em Remark 2}: Since $H$ is real, under complex conjugation it is 
$T(\epsilon )^* = T(\epsilon^*)$. Then $\xi_a (\epsilon )=\xi_a(\epsilon^*)$. 
The symplectic property (\ref{sigma}) with $B_n=I_m$ implies that the 
exponents of $T(\epsilon )$ come in pairs $\pm\xi_a $ for any $\epsilon$.

{\em Remark 3}: Since the 
transposed matrix $H(z^n)^t$ coincides with $H(z^{-n})$, then
$\det [\epsilon I_{nm}-H(z^n)]$ is a polynomial of degree 
$m$ of the variable $(z^n+z^{-n})$. 

{\em Remark 4}: Since $H_b(ze^{i2\pi/n})\simeq H_b(z)$ 
($\simeq $ means similarity) and $H_b(z)^* = H_b(z^*)$, the 
following symmetry holds:
$H_b (e^{\xi+ i (\frac{2\pi}{n}-\varphi)} )$ $\simeq$ 
$H_b(e^{\xi+i\varphi})^*$.

\subsection{The Lyapunov spectrum}
The localization properties of Anderson's model are usually derived from
the spectrum of positive Lyapunov exponents $\gamma_1<\ldots <\gamma_m $ 
of $T(\epsilon)^\dagger T(\epsilon )$, with $\epsilon $ real.
Oseledec's theorem\cite{Crisanti93} guarantees that for large $n$ 
it {\em does not depend on the length} $n$, 
{\em and on the realization of disorder}. The most interesting exponent
for physics is $\gamma_1$, that controls conductance. It is also the most 
difficult one to study numerically, because of the larger 
ones\cite{Pichard81,Kramer81}.
Thorough investigations of the Lyapunov spectrum, its statistical 
properties and scaling, have been done in 2D\cite{Slevin04} 
and 3D\cite{Markos97}. The influence of b.c. was studied\cite{Cerovski07} 
with the corner parameters $z^{\pm n}$ of the present theory being both 
replaced by the same parameter $t\in[0,1]$. It was found that the 
critical values of $\gamma_1$ and of the disorder parameter $w_c$ are
$t-$dependent, while the critical exponent $\nu $ is not.

Analytic results for the Lyapunov spectrum are accessible in perturbation 
theory for the 2D strip\cite{Rutman94,Rutman96,Romer04}, where $n$ is large
and $m$ is finite. In such quasi-1D Anderson systems, 
{\sl the large n limit of the exponent spectrum of T coincides with the Lyapunov spectrum}. For this reason here I concentrate on the spectral features of 
the matrix $H(z^n)$ to which $T$ is linked by duality.
I'll then show in another section that the spectral identity (\ref{spectrum}) 
allows to evaluate the smallest exponent $\xi_{min}$, which converges to
$\gamma_1$ for large $n$.

\subsection{1D Anderson model, Hatano and Nelson.}
\begin{figure}
\begin{center}
\includegraphics[width=4cm]{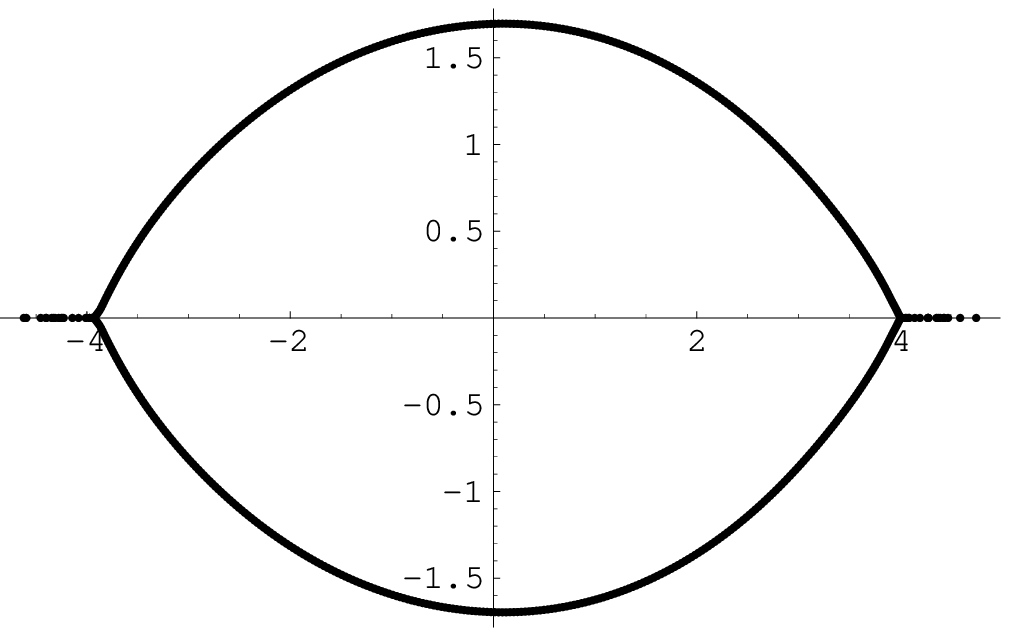}\hspace{0.5cm}
\includegraphics[width=4cm]{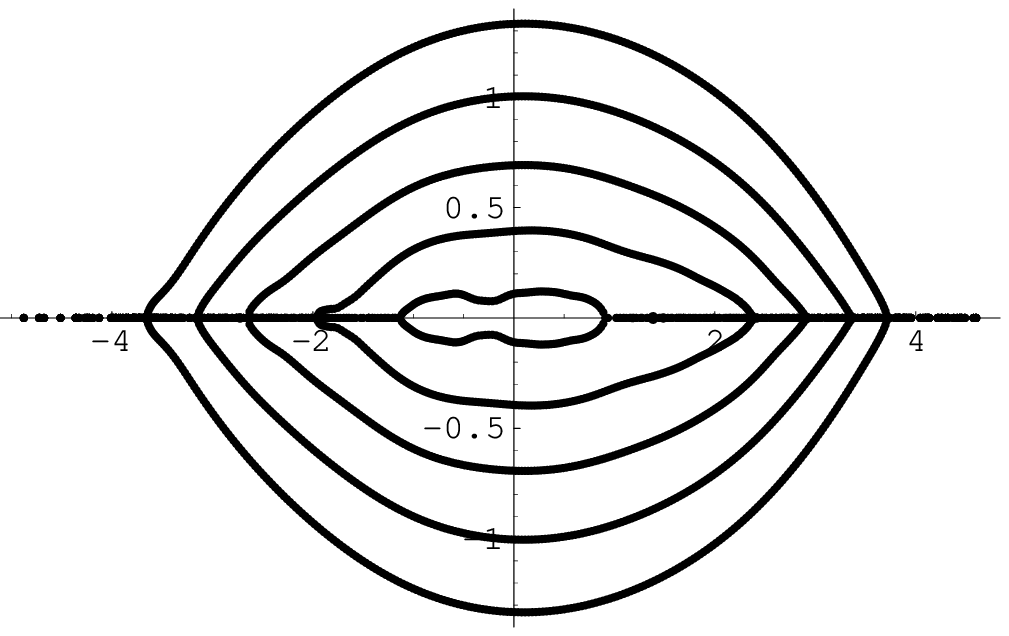}
\caption{\label{a01} 1D Anderson model: Left: the complex eigenvalues form 
a closed loop, with two wings of real values ($w=7$, $n=600$, $\xi =1$). 
Right: superposition of the eigenvalues of five matrices (same $w$ and $n$) 
for $\xi$ from $0.5$ (inner blob) to 1 (outer).}
\end{center}
\end{figure}

For a chain of $n$ sites the Hamiltonian is a tridiagonal $n\times n$ 
matrix and the transfer matrix is $2\times 2$. Hatano and Nelson\cite{Hatano96}
suggested to study 1D Anderson localization through the non-Hermitian 
extension of the model
\begin{eqnarray}
e^\xi \psi_{i+1}+e^{-\xi}\psi_{i-1} +v_i\psi_i = \epsilon \psi_i
\end{eqnarray}
with periodic b.c. As $\xi$ is increased from zero, the eigenvalues 
do not distribute randomly in the complex plane but form a loop, Fig.\ref{a01} 
(left), whose analytic expression is known for Cauchy 
disorder\cite{Goldsheid98}. 
The loop has two outer wings of real eigenvalues, that correspond to enough 
localized eigenstates, and evolves to a more and more regular shape, 
while the wings reduce, Fig.\ref{a01} (right).
The value $\xi_c (\epsilon )$ up to which an eigenvalue $\epsilon $ 
persists in the real axis, measures the inverse localization length of the 
physical eigenvector (the Lyapunov exponent): 
$\xi_c=\gamma_1 (\epsilon )$. The latter is evaluated through Herbert, Jones 
and Thouless' formula, 
\begin{equation} 
\gamma_1 (\epsilon ) = \int d\epsilon ' \rho (\epsilon') 
\log |\epsilon-\epsilon '|
\end{equation}
where $\rho (\epsilon )$ is the disorder-averaged level density of the matrix 
ensemble in the limit of large $n$, $\xi=0$.
The model has been studied by several 
authors\cite{Brouwer97,Shnerb98,Feinberg99,Janik99,Heinrichs01};
mathematical proofs were established by Goldsheid and 
Khoruzhenko\cite{Goldsheid03}. Diagonalization of large non-Hermitian
matrices is a delicate issue, as approximate eigenvalues may occur 
which are not close to true ones\cite{Davies01}.  

The Hatano-Nelson model is a case $m=1$ of the theory presented in 
Section II. The duality relation (\ref{duality}) simplifies greatly:
\begin{equation}
\det[\epsilon I_n -H(z^n)]= {\rm tr} T(\epsilon)  
- (z^n+z^{-n})\label{dualitym1}
\end{equation}
and implies that 
\begin{equation}
{\rm tr}T(\epsilon) =\det[\epsilon I_n-H(i)]\equiv p_n(\epsilon) \label{D}
\end{equation}
For a pure Bloch phase eq.(\ref{dualitym1}) describes the energy bands 
of $H(e^{in\varphi})$ as intersections of the polynomial
$y=p_n(\epsilon)$ with the strip $y=2\cos(n\varphi)$. As the non-Hermitian
regime is entered, $y=\pm2\cosh(n\xi)$, all eigenvalues of $H(e^{n\xi})$
are in the gaps, and approach pairwise for increasing $\xi$.
A pair collides at a zero of $p_n^\prime$ and becomes complex 
conjugate. This means that $2\cosh(n\xi)$ equals the height 
$|p_n|$ at an extremum of the polynomial.

\begin{figure}
\begin{center}
\includegraphics[width=4cm]{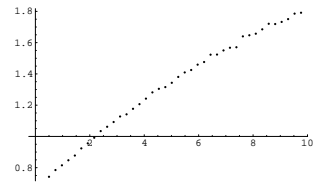}
\caption{\label{lyap} 1D Anderson model: the exponent $\xi_1(0)$ versus
disorder parameter $w$, for $n=600$, averaged over 8 samples of disorder.}
\end{center}
\end{figure}

The sum of the two eigenvalues $e^{\pm n(\xi_1+i\varphi_1)}$ of the transfer 
matrix is
\begin{equation}
p_n(\epsilon)=
\cosh(n\xi_1)\cos(n\varphi_1)+2i \sinh(n\xi_1)\sin(n\varphi_1)\nonumber
\end{equation}
Elimination of the phase results in an exact equation for 
the exponent $\xi_1(\epsilon )$
\begin{eqnarray}
\frac{({\rm Re}\,p_n)^2}{4\cosh^2 (n\xi_1)}+
\frac{({\rm Im}\,p_n)^2}{4\sinh^2 (n\xi_1)}=1
\end{eqnarray}
For large $n$ it becomes  
$e^{n\xi_1(\epsilon)} = |p_n(\epsilon )|$, and gives a convenient formula 
to compute the exponent (Fig.\ref{lyap})
\begin{equation}
\xi_1(\epsilon) = \frac{1}{n}\log |\det [\epsilon I_n -H(i)]|
\end{equation}
In 1D the single exponent is also given by the exact formula 
eq.(\ref{sumofexps}).
%
For large $n$, $\xi_1$ coincides with the Lyapunov exponent $\gamma_1$.

{\bf Proposition:} the eigenvalues of $H(e^{n\xi})$
distribute along the curve $\xi_1(\epsilon )=\xi$. Real eigenvalues 
(wings) solve $p_n(\epsilon )=2\cosh (n\xi)$. For large $n\xi$ the 
eigenvalues form the {\em Lemniscate}\cite{Hille} $|p_n(\epsilon)|=e^{n\xi}$.
\subsection{2D Anderson model}
\begin{figure}
\begin{center}
\includegraphics[width=3.5cm]{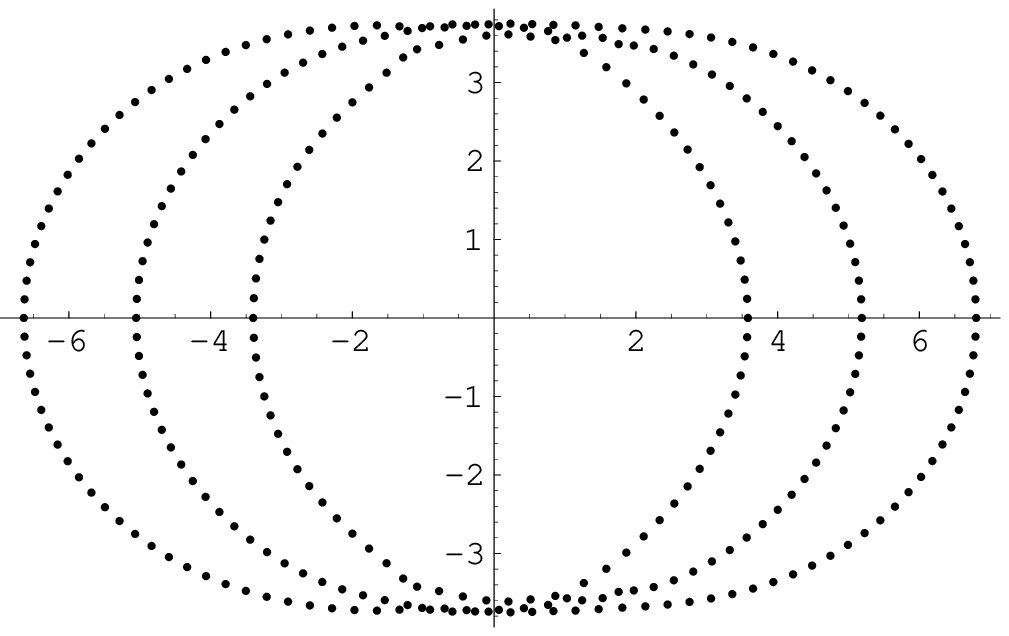}
\hspace{0.5cm}
\includegraphics[width=3.5cm]{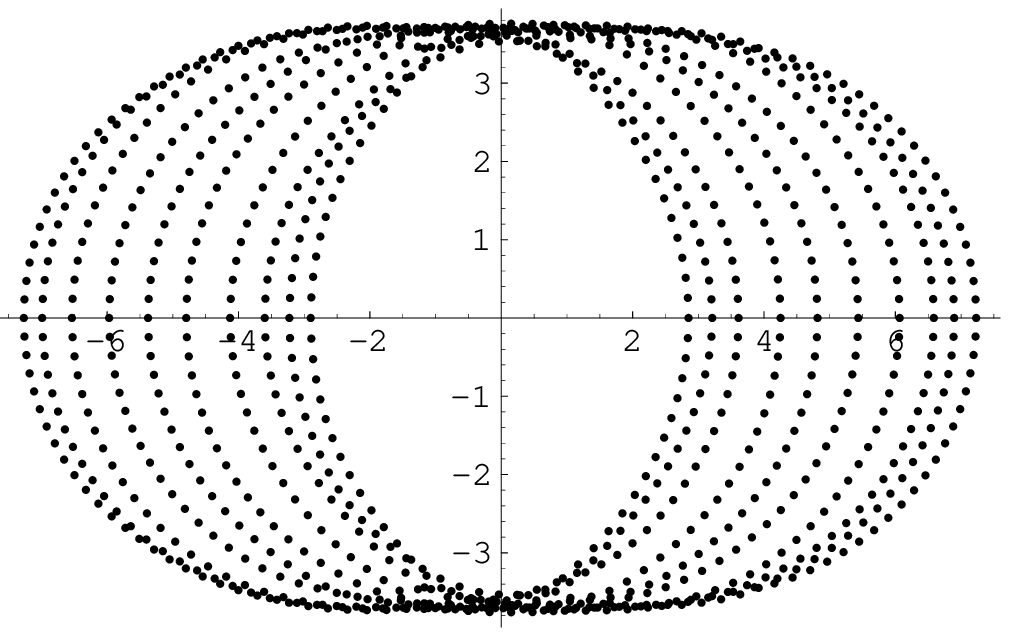}
\caption{\label{2Dstatic}2D Anderson model: eigenvalues of a single matrix, 
with parameters $w$=7, $n$=100, $\xi=1.5$, $\varphi =0$. Size of 
blocks: $m=3$ and $m=10$. The size $m$ of blocks is the number of loops.}
\end{center}
\end{figure}

For a rectangular $n\times m$ lattice the Hamiltonian matrix 
(\ref{andersonmatrix}) has diagonal blocks 
\begin{displaymath}
A_i =\left [\begin{array}{cccc}
v_{i,1} & 1   & {}    & 1      \\
1    & \ddots &\ddots & {}     \\
{}   & \ddots &\ddots & 1      \\
1    &   {}   & 1     & v_{i,m}\\
\end{array}\right ]
\end{displaymath}
The eigenvalue spectrum of 2D non-Hermitian Anderson model is studied, and 
explained in the light of duality. 
Fig.\ref{2Dstatic} shows the eigenvalues of two matrices $H(e^{n\xi})$
with same $\xi$ and different $m$. They are distributed along a number of 
loops which is precisely given by $m$, the size of the blocks. 

\begin{figure}
\begin{center}
\includegraphics[width=4cm]{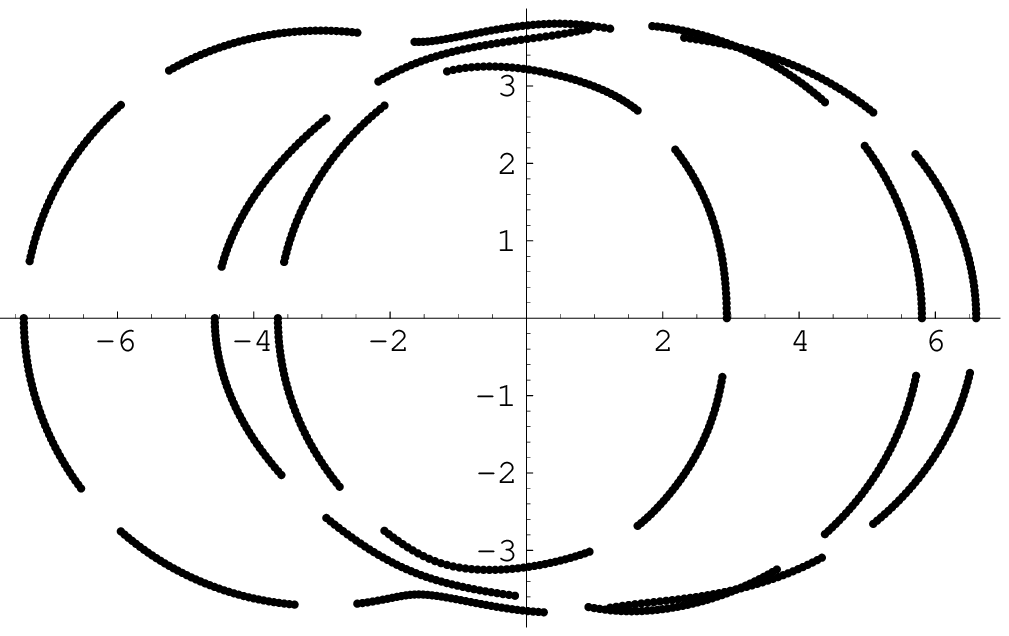}\hspace{0.5cm}
\includegraphics[width=4cm]{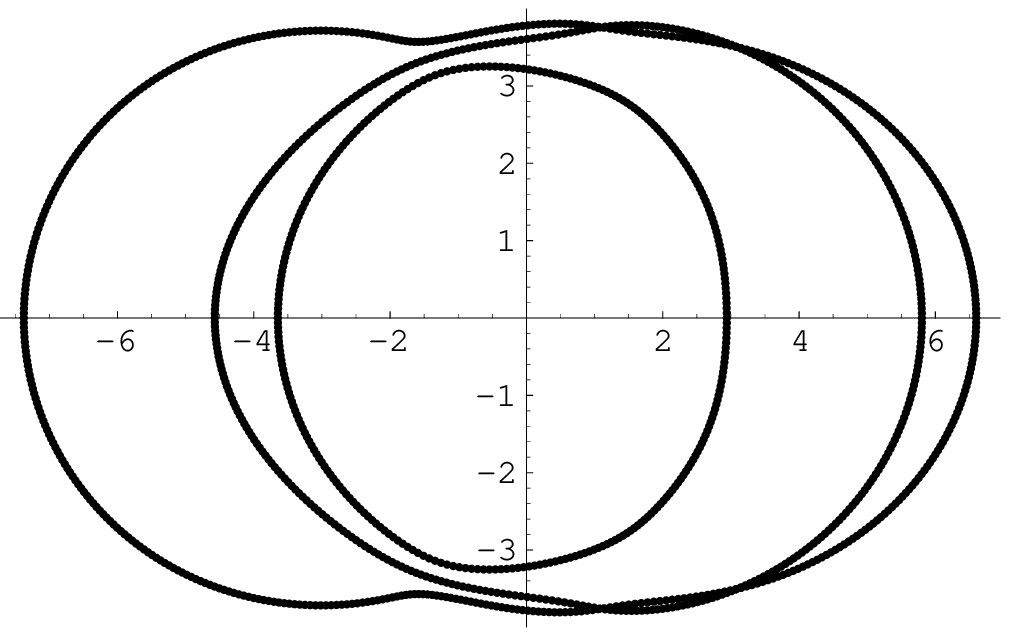}
\caption{\label{2Dphi} 2D Anderson model: motion of the eigenvalues in the 
complex plane for fixed disorder
w=7 and parameters $m$=3, $n$=8, and $\xi =1.5$ and varying the phase
$\varphi $. The
24 eigenvalues trace arcs:  
$0\le\varphi\le \pi/4-\delta$ (left) and $0\le\varphi\le \pi/4$ (right). 
The arcs join to form three loops. Loops are seen to contain different numbers 
of eigenvalues.}
\end{center}
\end{figure}

By varying only the phase $\varphi $ of $z$, the eigenvalues of the matrix
$H_b(z)$ move in the complex plane along arcs which retrace the loops. 
Fig.\ref{2Dphi} shows that, as $\varphi $ goes from $0$ to $2\pi/n $, an 
eigenvalue moves along an arc that terminates where the arc of another 
eigenvalue starts. The union of such consecutive arcs makes a loop, and
there are $m$ closed loops. 
Differently from the $w=0$ case, loops may contain different numbers 
of eigenvalues of the matrix.
The occurrence of loops is suggested by the duality equation: when a zero of 
$\det[H_b(z)-\epsilon I_{nm}]$ occurs, it is also a zero of 
$\det [T(\epsilon )-z^n]$, or
\begin{eqnarray}
\xi_a(\epsilon ) = \xi, \quad \varphi_a(\epsilon ) = \varphi 
\quad {\rm mod}\frac{2\pi}{n},\quad (a=1,\ldots , m) 
\end{eqnarray}
The loops are thus {\em level curves } of the exponents $\xi_a $, as functions 
of the complex variable $\epsilon $. 
In the limit of large $n$, the eigenvalues of one matrix fill $m$ loops.

\begin{figure}
\begin{center}
\includegraphics[width=4cm]{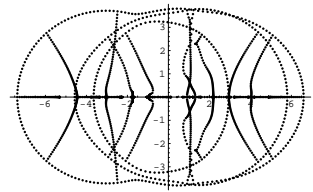}
\caption{\label{2Dxi}2D Anderson model. Motion of 24 eigenvalues for 
$0\le\xi\le 1.5$, $w$=7, $m=3$, $n$=8, $\varphi =0$. The various wings 
terminate on $m=3$ loops (at $\xi=1.5$ the phase is allowed to vary over
$2\pi$.)} 
\end{center}
\end{figure}

\begin{figure}
\begin{center}
\includegraphics[width=4cm]{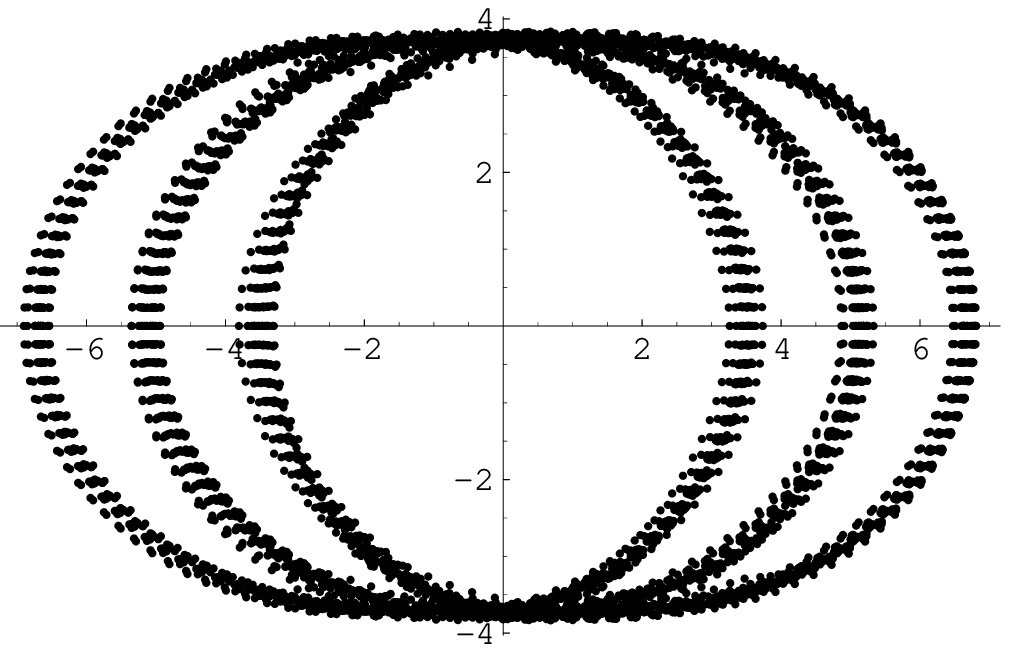}\hspace{0.2cm}
\includegraphics[width=4cm]{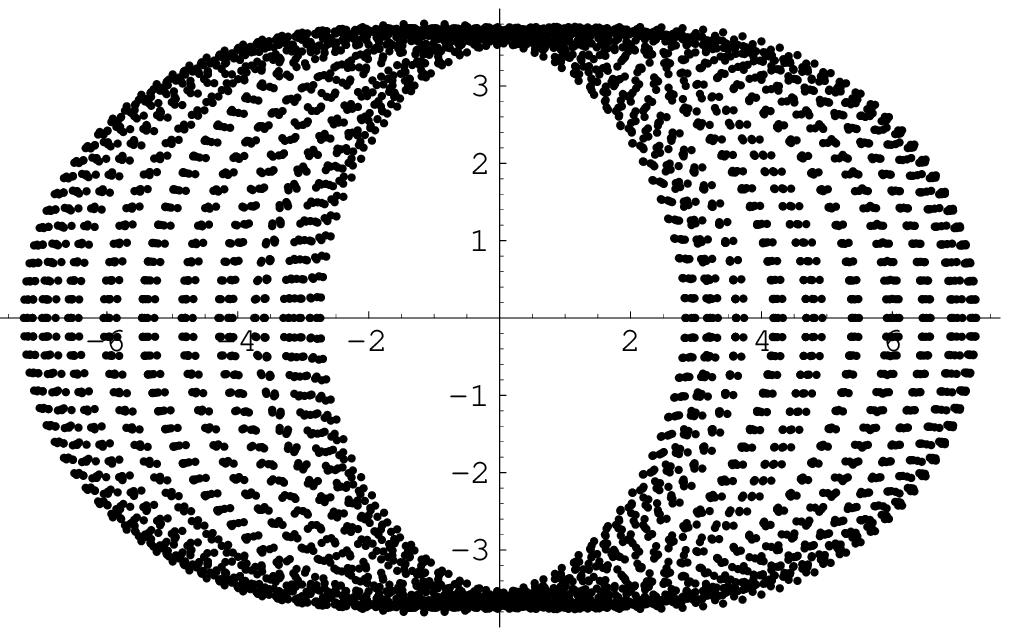}
\caption{\label{2Ddis} 2D Anderson model: superposition of eigenvalue spectra
for various realizations of disorder. For all: w=7, $n=100$, $\xi=1.5$, 
$\varphi =0$. Left: $m$=3, 20 realizations of disorder. 
Right: $m$=10, 5 realizations. The imaginary part of the eigenvalues is much 
less sensitive to disorder sampling than the real part.}
\end{center}
\end{figure}

%
%
%
In Fig.\ref{2Dxi} only $\xi$ is varied: the eigenvalues trace lines that
originate on the real axis (at $\xi=0$ the matrix is Hermitian). For zero 
disorder the lines would be arcs of hyperbola.
In the disordered case, for small $\xi $ the pattern of eigenvalues is 
complex, but evolves to regular loops. In the next section it is shown
that the evaluation of exponents requires a ``regular'' regime.
In Fig.\ref{2Ddis} the parameters $\xi $ and $\varphi $ are kept fixed, and
the eigenvalues are computed for different realizations of disorder, with same
strength $w$. They distribute along $m$ loops, that appear shifted along
the real axis for the different samples.

\subsection{The smallest exponent}
\begin{figure}
\begin{center}
\includegraphics[width=4cm]{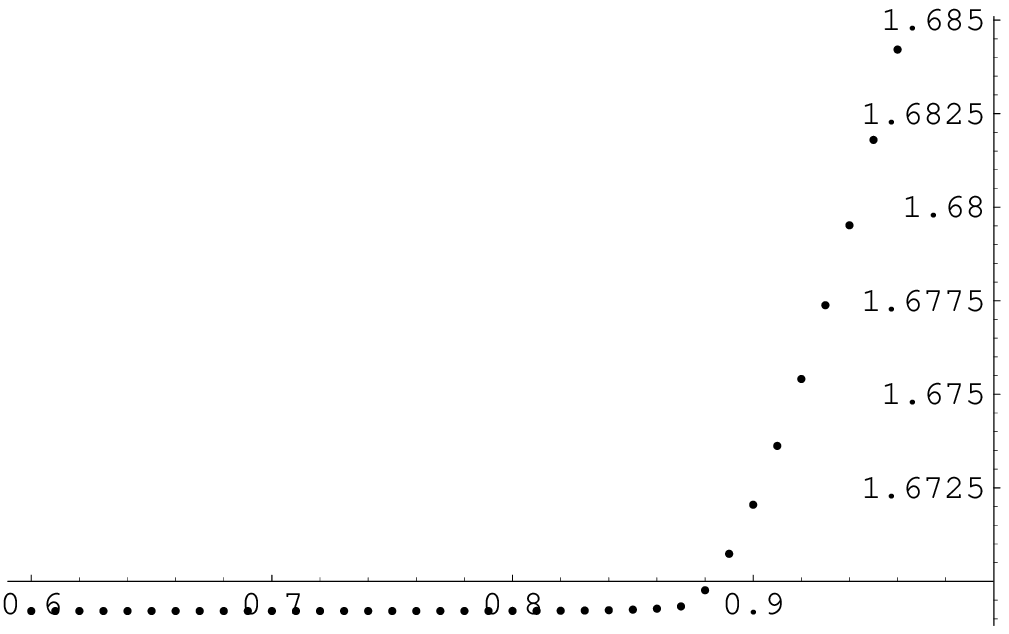}\hspace{0.5cm}
\includegraphics[width=4cm]{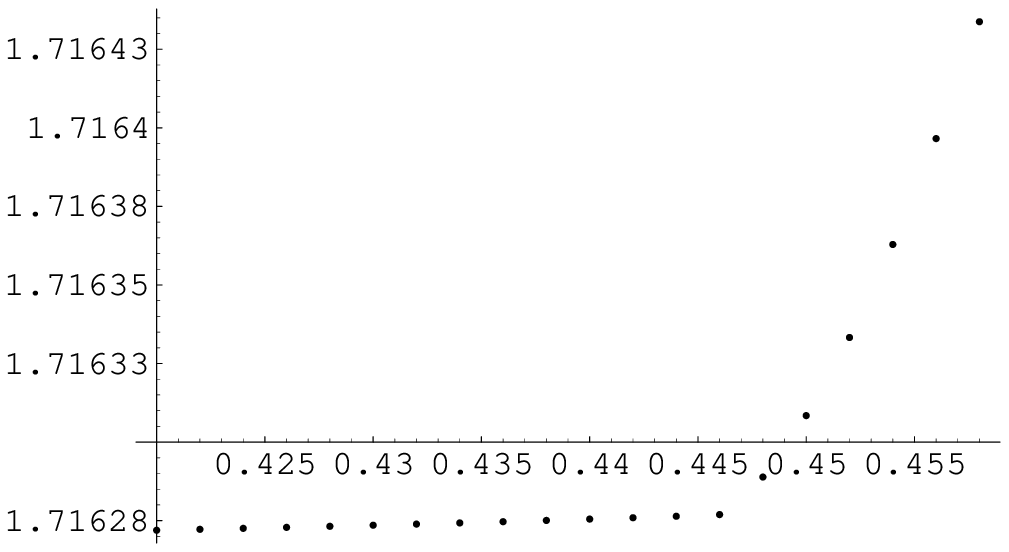}
\caption{\label{slope} 2D Anderson model: evaluation of the r.h.s. 
of eq.(\ref{spectrum}) as a function of $\xi<1$, for $\epsilon=0$, $w=7$, 
average on 40 angles. The change of slope marks $\xi_{min}(0)$. 
Left: $m=3$, $n=50$, $\xi_{min}(0)\approx 0.87$; the value 
1.6692 is the average $\frac{1}{3}(\xi_1+\xi_2+\xi_3)$.
Right: $m=n=20$,  $\xi_{min}(0)\approx 0.447$. The value 
1.71627 is the average $\frac{1}{20}(\xi_1+\ldots+\xi_{20})$. }
\end{center}
\end{figure}

\begin{figure}
\begin{center}
\includegraphics[width=5cm]{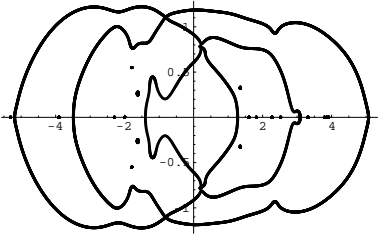}
\caption{\label{eigatslope} 2D Anderson model: the eigenvalues
for $\xi =0.835$, $\epsilon=0$, $w=7$, $m=3$, 
$n=50$. For $\xi $ near $\xi_{min}\approx 0.87$ the inner loop is isolated.}
\end{center}
\end{figure}

Let us assume that the exponents $\pm \xi_a$ of $T(\epsilon )$ are isolated, 
$0<\xi_{min}<\xi_2 \ldots <\xi_{m-1}<\xi_{MAX}$. By increasing $\xi $
from the value zero, the r.h.s. of eq. (\ref{spectrum}) 
yields a constant value (the $\xi=0$ value) until the value 
$\xi =\xi_{min}(\epsilon)$ is reached. Then the function becomes linear 
with slope $1/m$ until the value $\xi_2(\epsilon)$ is reached, where a new
change of slope occurs.
The change of slope can be used to identify the smallest exponent.
Fig.\ref{slope} illustrates this behaviour. Fig.\ref{eigatslope} shows 
that for $\xi\approx\xi_{min}$ the eigenvalues of the matrix are all well 
in the complex plane.

Kuwae and Taniguchi\cite{Kuwae01} extended Hatano Nelson's approach to 2D
Anderson model, and evaluated numerically the average critical value 
$\xi_c(w)$ where the first pair of eigenvalues turns to complex. They 
conjectured that the inverse of the localization length 
coincides with this critical value, i.e. $\gamma_1=\xi_c$. 
For example, for $m=n=20$, $w=7$ they evaluate $\xi_c\approx 0.1$. 
Fig.\ref{slope} shows that $\xi_{min}\approx 0.446$, a measure of
the inverse localization length. For this $\xi $ the spectrum is
evaluated and found in the complex plane. 
These facts adverse Kuwae's hypothesis 
that $\gamma_1$ signals the first escape to complex 
of the eigenvalues of Anderson's non-Hermitian Hamiltonian, and require 
further study.

\section{Conclusions}
Based on a spectral duality relation for block tridiagonal matrices
and Jensen's identity, the distribution of exponents of a transfer matrix 
can be evaluated from the eigenvalue spectrum of the Hamiltonian with 
non-Hermitian boundary conditions. 
A preliminary numerical study of the complex energy spectra of Anderson 
non-Hermitian Hamiltonian matrices is made. The spectra have support on 
loops, that are explained as sections of the exponents at fixed height: 
$\xi_a (\epsilon) =\xi$. This picture is 
complementary to a standard direct evaluation of the exponents 
$\xi_a (\epsilon)$ at fixed energy $\epsilon $, by diagonalization of the 
transfer matrix.

The spectral formula for exponents allows to evaluate the smallest one,
and involves eigenvalues of the non-Hermitian Hamiltonian that are
away from the real axis and distributed in loops that are nearly
untertwined. 
This, and the higher exponents, are interesting subjects for further 
investigation.
\vskip0.5truecm

\centerline{Appendix A}
We show the relationship of $Q(\epsilon )=T(\epsilon^*)^\dagger T(\epsilon )$ 
with a transfer matrix. It is convenient to factor $T(\epsilon)$ as
\begin{equation}
\left[\begin{array}{cc}B_n^{-1}&0\\ 0 & I_m\end{array}\right ]
t_n\sigma_{n-1} \cdots \sigma_1 t_1
\left[\begin{array}{cc}I_m &0\\ 0 & B_n^\dagger\end{array}\right ]
\label{factor}
\end{equation} 
Accordingly:
\begin{eqnarray}
&&Q(\epsilon )=(-1)^n 
\left[\begin{array}{cc}I_m &0\\ 0 & B_n\end{array}\right ]
u_1\sigma_1^\dagger \cdots \sigma_{n-1}^\dagger u_n\nonumber \\
&&\quad\times
\left[\begin{array}{cc}(B_n B_n^\dagger)^{-1} &0\\ 0 & I_m\end{array}\right ]
t_n \sigma_{n-1}\cdots \sigma_1 t_1
\left[\begin{array}{cc}I_m &0\\ 0 & B_n^\dagger\end{array}\right ]\nonumber
\end{eqnarray}
\begin{displaymath}
t_k =\left[\begin{array}{cc}\epsilon I_m- A_k & -I_m\\ 
I_m & 0\end{array}\right ],\quad
u_k =\left[\begin{array}{cc}A_k-\epsilon I_m& -I_m\\ 
I_m & 0\end{array}\right ],
\end{displaymath}
\begin{displaymath}
\sigma_k =
\left[\begin{array}{cc}B_k^{-1} &0\\ 0 & B_k^\dagger\end{array}\right ]
\end{displaymath}
To obtain the structure (\ref{factor}) of a transfer matrix, it is necessary
that $B_n^\dagger B_n=I_m$. The first and last factors containing $B_n$
are not consistent with (\ref{factor}), and only the intermediate product
$u_1\sigma_1^\dagger \cdots \sigma_1t_1 $
is the transfer matrix of a tridiagonal block matrix, eq.(\ref{teta}). 
\vskip0.5truecm

\centerline{Appendix B}
One can show the following properties of the matrix $K(s)$.
$J$ is the matrix with $2n$ blocks $I_m$ along the diagonal from
lower left to upper right corners; $S_3$ is the block diagonal matrix 
$\{I_{nm},-I_{nm}\}$. Then:
\begin{eqnarray}
JK(s)J = K(s^*)^\dagger\\
S_3 K(s)S_3 =K(1/s^*)
\end{eqnarray}
The two imply that $K(s)$ is similar to $K(1/s)$.
\vskip0.5truecm

\centerline{Appendix C}
For the Anderson model with no disorder ($w=0$), real energy $\epsilon$ 
and large $n$, the exponents of $T$ and $T^\dagger T$ coincide. Proof: It is  
\begin{eqnarray} 
&&T=
\left[\begin{array}{cc} \epsilon I_m -A & -I_m\\ I_m & 0\end{array}\right ]^n\\
&&=\left[\begin{array}{cc} U & 0\\ 0 & U\end{array}\right ] 
\left[\begin{array}{cc} \epsilon I_m
-\Lambda & -I_m\\ I_m & 0\end{array}\right ]^n
\left[\begin{array}{cc} U^\dagger & 0\\ 0 & U^\dagger\end{array}\right ]   
\nonumber
\end{eqnarray}
where $A=U\Lambda U^\dagger $ and $\Lambda $ is the diagonal matrix of 
eigenvalues $\{\lambda_1,\ldots \lambda_m\}$. 
The eigenvalues of $T$ are $m$ pairs $z_k^{\pm n}$, where $z_k$
is a root of the equation $z_k^2-(\epsilon -\lambda_k)z_k+1=0$. 

The power $n$ of the matrix can be computed by means of Cayley-Hamilton's 
formula. Because the blocks are diagonal, only powers zero and one of the 
matrix are needed:
\begin{eqnarray}
\left[\begin{array}{cc} \epsilon I_m-\Lambda & -I_m\\ I_m & 0\end{array}\right ]^n =
\left[\begin{array}{cc} \alpha & 0\\ 0 & \alpha \end{array}\right ] +
\left[\begin{array}{cc} \beta  & 0\\ 0 & \beta  \end{array}\right ]
\left[\begin{array}{cc} \epsilon I_m-\Lambda & -I_m\\ I_m & 0\end{array}\right ]
\nonumber
\end{eqnarray}
$\alpha $ and $\beta $ are diagonal matrices with elements constructed with
the roots $z_k$: $z_k^{\pm n} = \alpha_k+\beta_k z_k^{\pm 1}$. Since $\epsilon $ is real, $\alpha_k$ and $\beta_k$ are real.
The matrix $T^\dagger T$ is then constructed, and diagonalized. Its
eigenvalues are pairs $w_k^{\pm 1}$, with sum
\begin{eqnarray}
w_k+w_k^{-1} = (z_k^{2n}+z_k^{-2n})\left (\frac{z_k^2+1}{z_k^2-1}\right )^2
-\frac{8z_k^2}{(z_k^2-1)^2}
\end{eqnarray}
If $|z_k|>1$ then, for large $n$, 
$|w_k|\approx |z_k|^{2n}$ (the spectrum of Lyapunov exponents of 
$T^\dagger T$ and the spectrum of exponents of $T$ coincide)$\blacksquare $
\vskip 0.5truecm

{\bf Acknowledgements}
I wish to thank professors Leonid Pastur, Borys Khoruzhenko, and especially 
Ilya Goldsheid for their interesting suggestions, while at the meeting
``Anderson Localization and Related Phenomena'' (aug 2008, Newton 
Institute, Cambridge).

\vfill
\end{document}